\documentclass[english,british]{llncs}
\usepackage{lmodern}
\usepackage[T1]{fontenc}
\usepackage[latin9]{inputenc}
\usepackage{amsmath}
\usepackage{graphicx}

\makeatletter
\@ifundefined{showcaptionsetup}{}{%
 \PassOptionsToPackage{caption=false}{subfig}}
\usepackage{subfig}
\makeatother

\graphicspath{{././}}

\usepackage{babel}
\begin{document}

\title{Hierarchical Dirichlet Process for Tracking Complex Topical Structure Evolution and Its Application to Autism Research Literature}

\author{Adham Beykikhoshk, Ognjen~Arandjelovi\'c, Dinh~Phung, Svetha~Venkatesh}
\institute{Deakin University, Australia}

\maketitle
\vspace{-10pt}
\begin{abstract}
In this paper we describe a novel framework for the discovery of the topical content of a data corpus, and the tracking of its complex structural changes across the temporal dimension. In contrast to previous work our model does not impose a prior on the rate at which documents are added to the corpus nor does it adopt the Markovian assumption which overly restricts the type of changes that the model can capture. Our key technical contribution is a framework based on (i) discretization of time into epochs, (ii) epoch-wise topic discovery using a hierarchical Dirichlet process-based model, and (iii) a temporal similarity graph which allows for the modelling of complex topic changes: emergence and disappearance, evolution, and splitting and merging. The power of the proposed framework is demonstrated on the medical literature corpus concerned with the autism spectrum disorder (ASD) -- an increasingly important research subject of significant social and healthcare importance. In addition to the collected ASD literature corpus which we will make freely available, our contributions also include two free online tools we built as aids to ASD researchers. These can be used for semantically meaningful navigation and searching, as well as knowledge discovery from this large and rapidly growing corpus of literature.
\end{abstract}

\vspace{-10pt}\section{Introduction\label{sec:Introduction}}\vspace{-4.5pt}
The autism spectrum disorder (ASD) is a life-long neurodevelopmental disorder with poorly understood causes on the one hand, and a wide range of potential treatments supported by little evidence on the other. The disorder is characterized by severe impairments in social interaction, communication, and in some cases cognitive abilities. Considering the social and economic burden of ASD it is unsurprising that it has been attracting an increasing amount of research attention which has resulted in a rapid growth of the relevant corpus of literature. Navigating this vast amount of data by conventional, manual means is difficult and limiting; yet the rapid rise in the diagnosis rate of ASD demands timely research on its aetiology and treatment. Consequently, the potential benefit of tools based on novel data-mining and machine learning techniques is immense~\cite{BeykAranPhunVenk+2014}. More meaningful ways for visualising or searching for data could provide invaluable information in clinical and administrative decision making as well as aid research, while automatic knowledge discovery would in its own right advance the understanding of the underlying  phenomena (e.g.\ epidemiological patterns). We describe a novel method which contributes towards this goal.

More specifically, we describe a general framework for the analysis of medical literature capable of (i) discovering the underlying topical structure, (ii) inferring the relationships between different discovered topics, and (iii) tracking the evolution of topics over time. The proposed framework uses the hierarchical Dirichlet process (HDP) to extract topics automatically, and then constructs a similarity graph over them using an inter-topic similarity measure; topic evolution over time can be inferred from this graph. The effectiveness of our approach is demonstrated on the specific example of a large longitudinal data corpus of medical literature on ASD which we collected. This corpus includes more than 18,000 articles published over the course of 42 years.  In addition to the aforementioned technical contributions, our further contribution is this corpus which will be made public following the publication of the present paper.


The results we report on the collected ASD literature corpus illustrate the usefulness of our method and its ability to extract and track over time abstract topical knowledge, inferring the point at which a certain topic comes into existence, how its evolves, splits into multiple new topics or merges with the existing ones, and lastly when it ceases to exist. This is demonstrated on examples of well-known research directions in the field, making out work  the first to examine the medical literature on ASD using advanced topic modelling tools. Our additional contributions come in the form of two free online tools which allow researchers to (i) navigate and search the literature in a semantically meaningful manner (see \url{http://www.undersdtanfigutism.tk}), and (ii) understand the development and relationships between different ideas which permeate research in the domain of ASD (see \url{http://www.goo.gl/Ws7V64}).

\vspace{-4.5pt}\section{Previous work\label{sec:Background}}\vspace{-4.5pt}
In this section we review the most relevant previous work on topic modelling. We focus our attention first on latent topic models which have dominated the field in the last decade, and then on biomedical text mining, given the application domain within which our framework is evaluated in Section~\ref{sec:Experiments}.

\vspace{-4.5pt}\subsection{Latent topic models\label{sub:Latent-Topic-Models}}\vspace{-4.5pt}
An important early topic modelling approach is the latent semantic indexing (LSI)~\cite{deerwester1990indexing} which remains popular. Two notable limitations of LSI are its inability to deal effectively with polysemy and to produce an explicit description of the latent space. A probabilistic improvement overcomes these by explicitly characterizing the latent space with semantic topics, and by employing a probabilistic generative model that addresses the polysemy problem~\cite{hofmann1999probabilistic}. Nevertheless, probabilistic LSI is prone to parameter overfitting caused by an uncontrolled growth in the number of parameters as the document corpus is increased. In addition, the necessary assignment of probabilities to documents is a nontrivial task~\cite{blei2003latent}.

The recently proposed latent Dirichlet allocation (LDA) method~\cite{blei2003latent} overcomes the overfitting problem by adopting a Bayesian framework and a generative process at the document level. While LDA has quickly become a standard tool for topic modelling, it too experiences challenges when applied on real-world data. In particular, being a parametric model the number of desired output topics has to be specified in advance. The HDP model as the nonparametric counterpart of LDA was introduced by Teh \textit{et al.}~\cite{teh_etal_2006_hierarchical} and addressed this limitation by using a Dirichlet process (DP) (as opposed to a Dirichlet distribution) as the prior on topics. Therefore, each document is modelled using an infinite mixture model, allowing the data to inform the complexity of the model and infer the number of resulting topics automatically. We discuss this model in further detail in Section~\ref{sec:Framework}.

\vspace{-4.5pt}\subsubsection{Temporal topic modelling}\vspace{-4.5pt}
A notable limitation of most models described in the existing literature lies in their assumption that the data corpus is static; this includes those based on LDA mentioned previously, or the hierarchical Dirichlet process described in detail in the next section. However, in many practical applications documents are added to the corpus in a temporal manner and their ordering has significance (non-exchangeability property).  As a consequence, the topical structure of the corpus changes over time. The assumption made by all previous work, and indeed adopted by us, is that documents are not exchangeable at large temporal scales but are at short time scales, thus treating the corpus at temporally locally static.

The existing work on temporal topic modelling can be divided into two groups of approaches both of which can be based on parametric~\cite{blei2006dynamic,Wang2008,Wang2006_tot} or nonparametric~\cite{ren2008dynamic,zhang2010evolutionary} techniques, the former suffering from the limitation that they contain free parameters which must be set \textit{a priori}. Methods of the first group discretize time into epochs, apply a static topic model to each epoch, and by making the Markovian assumption relate the parameters of each epoch's topic model to those of the epochs adjacent to it in time~\cite{blei2006dynamic,Wang2008,ren2008dynamic,zhang2010evolutionary}. While the approach we propose in this paper adopts the idea of time discretization, it diverges in its other features from this group of methods thereafter. In particular, instead of employing the Markovian assumption we describe a novel structure in form of a temporal similarity graph, which gives our method greater flexibility, as described in detail in the next section. The second group of methods in the literature regard document time-stamps as observations of a continuous random variable~\cite{Wang2006_tot,dubey2013nonparametric}. This assumption severely limits the type of topic changes which can be described. For example, as opposed to our model, these models are not capable of describing the evolution of topics, or their splitting and merging, and are rather constrained to tracking simple topic popularity (rise/fall).

\vspace{-4.5pt}\subsection{Biomedical text mining}\vspace{-4.5pt}
Most previous work on text-based knowledge discovery has rather focused on (i) the tagging of names of entities such as genes, proteins, and diseases~\cite{settles2005abner}, (ii) the discovery of relationships between different entities e.g.\ functional associations between genes~\cite{rhodes2004oncomine}, or (iii) the extraction of information pertaining to events such as gene expression or protein binding~\cite{simpson2012biomedical}.

The idea that the medical literature could be mined for new knowledge is typically attributed to Swanson~\cite{swan1986}. For example by manually examining medical literature databases he hypothesised that dietary fish oil could be beneficial for Raynaud's syndrome patients, which was later confirmed by experimental evidence. Work that followed sought to develop statistical methods which would make this process automatic. Most approaches adopted the use of term frequencies and co-occurrences using dictionaries such as Medical Subject Headings (MeSH)~\cite{Roge1963}.

Most existing work on biomedical knowledge discovery is based on what may be described as traditional data mining techniques (neural networks, support vector machines etc); comprehensive surveys can be found in~\cite{kumar2014biomedical,simpson2012biomedical}. The application of state-of-the-art Bayesian methods in this domain is scarce. Amongst the notable exceptions is the work by Blei \textit{et al.}  who showed how latent Dirichlet allocation (LDA) can be used to facilitate the process of hypothesis generation in the context of genetics~\cite{blei2006statistical}. Arnold \textit{et al.} used a similar approach to demonstrate that abstract topic space representation is effective in patient-specific case retrieval~\cite{arnold2010clinical}. In their later work they introduced a temporal model which learns topic trends and showed that the inferred topics and their temporal patterns correlate with valid clinical events and their sequences~\cite{arnold2012topic}.  Wu \textit{et al.} used LDA for gene-drug relationship ranking~\cite{wu2012ranking}.

\vspace{-4.5pt}\section{Proposed framework\label{sec:Framework}}\vspace{-4.5pt}
We begin this section by reviewing the relevant theory underlying HDP mixture modelling which plays the central rule in the proposed framework. Then we turn our attention to the main technical contribution of our work and explain how the HDP is employed to discover the topical content of a literature corpus and track its structural changes over time.

\vspace{-4.5pt}\subsection{Hierarchical Dirichlet process mixture models\label{sub:Hieraqrchical-Dirichlet-Process}}\vspace{-4.5pt}
The Dirichlet process is a useful prior for mixture modelling which allows a document collection to accommodate a potentially infinite number of topics. It is the building block of Bayesian nonparametric methods. A Dirichlet process~\cite{ferguson1973bayesian} $\text{DP}\left(\gamma,H\right)$ is defined as a distribution of a random probability measure $G$ over a measure space $\left(\Theta,\mathcal{B},\mu\right)$, such that for any finite measurable partition $\left(A_{1},A_{2},\ldots,A_{r}\right)$ of $\Theta$ the random vector $\left(G\left(A_{1}\right),\ldots,G\left(A_{r}\right)\right)$
is a Dirichlet distribution with parameters $\left(\gamma H\left(A_{1}\right),\ldots,\gamma H\left(A_{r}\right)\right)$. An alternative view of the DP emerges from the so-called stick-breaking process which adopts a constructive approach using a sequence of discrete draws~\cite{sethuraman1991constructive}. Specifically, if $G\sim\text{DP}\left(\gamma,H\right)$ then $G=\sum_{k=1}^{\infty}\beta_{k}\delta_{\phi_{k}}$ where $\phi_{k}\overset{iid}{\sim}H$ and $\boldsymbol{\beta}=\left(\beta_{k}\right)_{k=1}^{\infty}$ is the vector of weights obtained as $\beta_{k}=v_{k}\prod_{l=1}^{k-1}\left(1-v_{l}\right)$ and $v_{l}\overset{iid}{\sim}\text{\text{Beta}}\left(1,\gamma\right)$.

Owing to the discrete nature and infinite dimensionality of its draws, the DP is a highly useful prior for Bayesian mixture models. By associating different mixture components with atoms $\phi_{k}$ of the stick-breaking process, and assuming $x_{i}|\phi_{k}\overset{iid}{\sim}F\left(x_{i}|\phi_{k}\right)$ where $F\left(.\right)$ is the likelihood kernel of the mixing components, we can formulate the Dirichlet process mixture model (DPM). The DPM is suitable for nonparametric clustering of exchangeable data in a single group e.g.\ words in a document where the DPM models the underlying structure of the document with potentially an infinite number of topics. However, many real-world problems are more appropriately modelled as comprising multiple groups of exchangeable data (e.g.\ a collection of documents). In such cases it is usually desirable to model the observations of different groups jointly, allowing them to share their generative clusters. This idea is known as the sharing statistical strength and is achieved using a hierarchical structure.

\begin{figure}[t]
\vspace{-15pt}
  \centering
  \subfloat[HDP\label{fig:HDP-graphical-model}]{\includegraphics[width=0.38\textwidth]{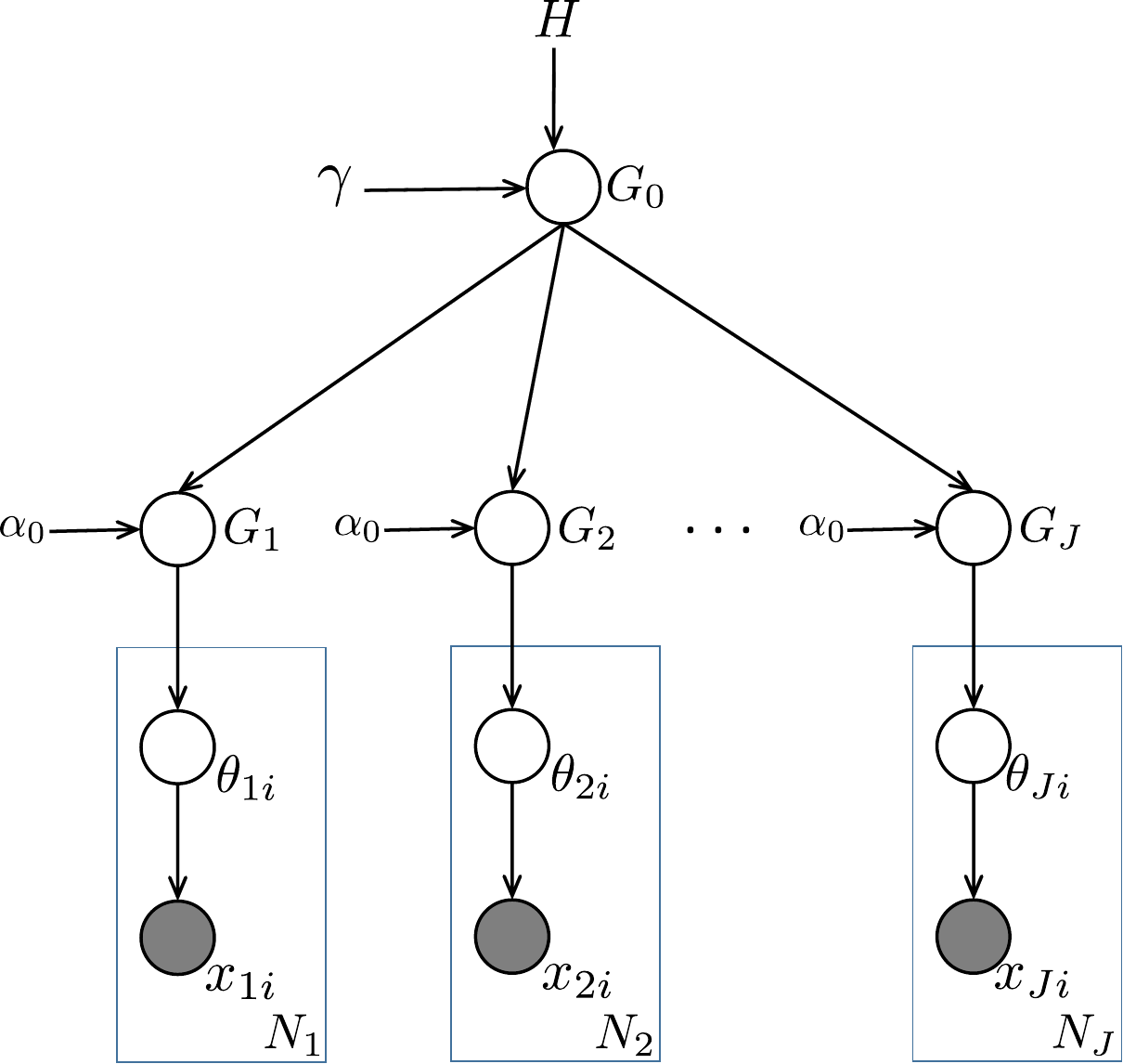}}
  \subfloat[Proposed\label{fig:Framework-graphical-model}]{\includegraphics[width=0.58\textwidth]{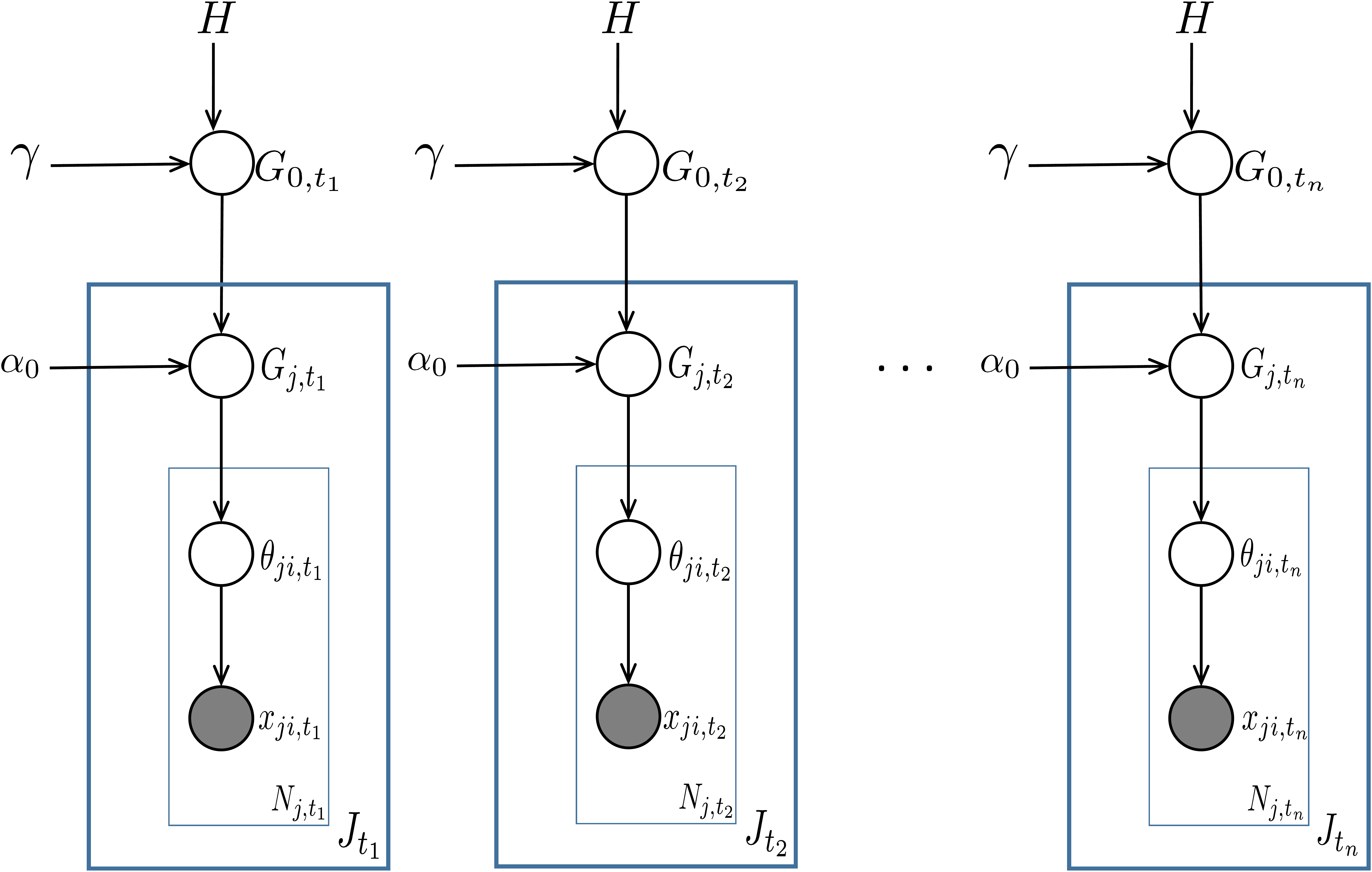}}
  \caption{(a) Graphical model representation of HDP. Each box represents one document whose observed data (words) is shown shaded. Unshaded nodes represent latent variables. An observed datum $x_{ji}$ is assigned to a latent mixture component parametrized by $\theta_{ji}$. $\gamma$ and $\alpha$ are the concentration parameters and $H$ is the corpus-level base measure. (b) Graphical model representation of the proposed framework. The corpus is temporally divided into $t_{n}$ epochs and each epoch modelled using an HDP (outer boxes).}
  \vspace{-12pt}
\end{figure}

Amongst different ways of linking group-level DPMs, HDP~\cite{teh_etal_2006_hierarchical} offers an interesting solution whereby base measures of document-level DPs are drawn from another DP. In this way the atoms of the corpus-level DP (i.e.\ topics in our case) are shared across the corpus. Formally, if $\mathbf{x}=\left\{ \mathbf{x}_{1},\ldots,\mathbf{x}_{J}\right\} $ is a document collection where $\mathbf{x}_{j}=\left\{ x_{j1},\ldots,x_{jN_{j}}\right\} $ is the $j$-th document comprising $N_{j}$ words, each document is modelled with a DPM $G_{j}|\alpha_{0},G_{0}\stackrel{iid}{\sim}\text{DP}\left(\alpha_{0},G_{0}\right)$ where its DP prior is further endowed by another DP $G_{0}|\gamma,H\sim\text{DP}\left(\gamma,H\right)$. This is illustrated schematically in Figure~\ref{fig:HDP-graphical-model}. Since the base measure of $G_{j}$ is drawn from $G_{0}$, it takes the same support as $G_{0}$. Also the parameters of the group-level mixture components, $\theta_{ji}$, share their values with the corpus-level DP support on $\left\{ \phi_{1},\phi_{2},\ldots\right\} $. Therefore $G_{j}$ can be equivalently expressed using the stick-breaking process as $G_{j}=\sum_{k=1}^{\infty}\pi_{jk}\delta_{\phi_{k}}$ where $\boldsymbol{\pi}_{j}|\alpha_{0},\gamma\sim\text{DP}\left(\alpha_{0},\gamma\right)$\cite{teh_etal_2006_hierarchical}. The posterior for $\theta_{ji}$ has been shown to follow a Chinese restaurant franchise process which can be used to develop inference algorithms based on Gibbs sampling~\cite{teh_etal_2006_hierarchical}.

\vspace{-4.5pt}\subsection{Modelling topic evolution over time \label{sub:Modelling-Topics-Sequentially}}\vspace{-4.5pt}
In this section we show how the described HDP-based model can be applied to the analysis of temporal topic changes in a longitudinal data corpus. We begin by dividing the literature corpus by time into multiple \emph{epochs}. Each epoch is then modelled separately using an HDP.  Different epochs' models are inferred using the same initial corpus-level base measure and hyperparameters. Hence if $n$ is the number of epochs, we obtain $n$ sets of  topics $\boldsymbol{\theta}=\left\{ \boldsymbol{\theta}_{t_{1}},\ldots,\boldsymbol{\theta}_{t_{n}}\right\}$ where $\boldsymbol{\theta}_{t}=\left\{ \theta_{1,t},\ldots,\theta_{K_{t},t}\right\} $ is the set of topics that describe epoch $t$, and $K_{t}$ their number (which is inferred automatically, as described previously). This is illustrated in Figure~\ref{fig:Framework-graphical-model}. In the next section we describe how given an inter-topic similarity measure the evolution of different topics across epochs can be tracked.

\vspace{-4.5pt}\subsection{Measuring topic similarity\label{sub:Modelling-Topics-Similarity}}\vspace{-4.5pt}
Our goal now is to track changes in the topical structure of a data corpus over time. The simplest changes of interest include the emergence of new topics, and the disappearance of others. More subtly, we are also interested in how a specific topic changes -- how it evolves over time in terms of the contributions of different words it comprises, as well as how it splits into new topics or merges with the existing ones. Clearly this information can provide valuable insight into the refinement of ideas and findings in the scientific community, effected by new research and accumulating evidence.

The key idea behind our approach stems from the observation that while topics may change significantly over time, by their very nature the change between successive epochs is limited. Therefore we infer the continuity of a topic in one epoch by relating it to all topics in the immediately subsequent epoch which are sufficiently similar to it under some similarity measure. This can be seen to lead naturally to a similarity graph representation whose nodes correspond to topics and whose edges link those topics in two epochs which are related. Formally, the weight of the directed edge that links $\phi_{j,t}$ , the $j$-th topic in epoch $t$, and $\phi_{k,t+1}$ is set equal to $\rho\left(\phi_{j,t},\phi_{k,t+1}\right)$ where $\rho$ is an appropriate similarity measure. Given that in our HDP-based model each topic is represented by a probability distribution, suitable similarity metrics include the Jaccard similarity, the Jenson-Shannon divergence, or the $L_2$-norm for example.

A conceptual illustration of a similarity graph is shown in Figure~\ref{fig:topic-similarity-graph}. It shows three consecutive time epochs $t-1,t,$ and $t+1$ and a selection of topics in these epochs. Graph edge weight i.e.\ inter-topic similarity is encoded by varying the thickness of the corresponding line connecting two nodes -- a thicker line signifies more similar topics. We use a threshold to eliminate automatically weak edges, retaining only the edges which correspond to sufficiently similar topics in adjacent epochs. It can be seen that this readily allows us to detect the disappearance of a particular topic, the emergence of new topics, as well as the splitting or merging of different topics:
\begin{list}{}{\leftmargin=20pt}
  \item[\bf Emergence] If a node does not have any edges incident to it, the corresponding topic is taken as having emerged in the associated epoch (e.g.\ $\phi_{j+2}$ at time $t$ in Figure~\ref{fig:topic-similarity-graph}).\\[-5pt]

  \item[\bf Disappearance] If no edges originate from a node, the corresponding topic is taken to vanish in the associated epoch (e.g. $\phi_{j}$ at time $t$ in Figure~\ref{fig:topic-similarity-graph}).\\[-5pt]

  \item[\bf Splitting] If more than a single edge originates from a node, the corresponding topic is understood as being split into multiple topics in the next epoch (e.g.\ $\phi_{i}$ is split into $\phi_{j}$ and $\phi_{j+1}$ in Figure~\ref{fig:topic-similarity-graph}).\\[-5pt]

  \item[\bf Merging] If more than a single edge is incident to a node, the topics of the nodes from which the edges originate are understood as having merged together to form a new topic (e.g.\ $\phi_{i}$ and $\phi_{i+1}$ merge to form $\phi_{j+1}$ in Figure~\ref{fig:topic-similarity-graph}).
\end{list}

\vspace{-4.5pt}\section{Experimental evaluation\label{sec:Experiments} }\vspace{-4.5pt}
Having introduced the main technical contribution of our work we now illustrate its usefulness on the example of ASD literature analysis, and describe additional contributions in the form of two free online tools that we developed to aid ASD researchers.

\vspace{-4.5pt}\subsection{Data collection\label{sub:Dataset}}\vspace{-4.5pt}
To the best of our knowledge there are no publicly available corpora of ASD-related medical literature. Hence we collected a comprehensive dataset ourselves, which will be made public following the acceptance of the present paper. We describe our collection methodology and the pre-processing of data we performed to extract standard  features used for text analysis.

\vspace{-4.5pt}\subsubsection{Raw data collection\label{sub:Data-acquisition}}\vspace{-4.5pt}
We used the PubMed search engine that allows users to access the US National Library of Medicine for abstracts and references of life science and biomedical scholarly articles. We assumed a paper is related to ASD if the term ``autism'' is present in its title or abstract, and collected only papers written in English. The earliest publication fitting our criteria is that by Kanner~\cite{kanner1946irrelevant}, and we collected all matching publications up to the final one indexed by PubMed on 24th July 2014, yielding a corpus of 20,138 publications. We discarded the 1,946 which do not have an abstract indexed, ending with the total of 18,192 papers in our dataset. We used the abstracts text to evaluate our method.

\vspace{-4.5pt}\subsubsection{Data pre-processing\label{sub:Data-pre-processing}}\vspace{-4.5pt}
Following the standard practice in text processing literature we applied soft lemmatization on the abstracts in our dataset, using the freely available WordNet tool~\cite{MillBeckFellGros+1990}. No stemming was performed to avoid potential distortion of words which is sometimes effected by heuristic rules used by stemming algorithms. After lemmatization and the removal of so-called stop words, we obtained 1.9 million terms in the entire corpus when repetitions are counted, and 37,278 unique terms. We construct the vocabulary for our method by selecting the subset of the most frequent unique terms which explain 90\% of the energy of the corpus, which resulted in a 3,738 term vocabulary.

\vspace{-4.5pt}\subsection{Proposed method implementation}
\begin{figure}
  \vspace{-10pt}
  \centering
  \subfloat[Topic similarity graph\label{fig:topic-similarity-graph}]
     {\includegraphics[width=0.5\textwidth]{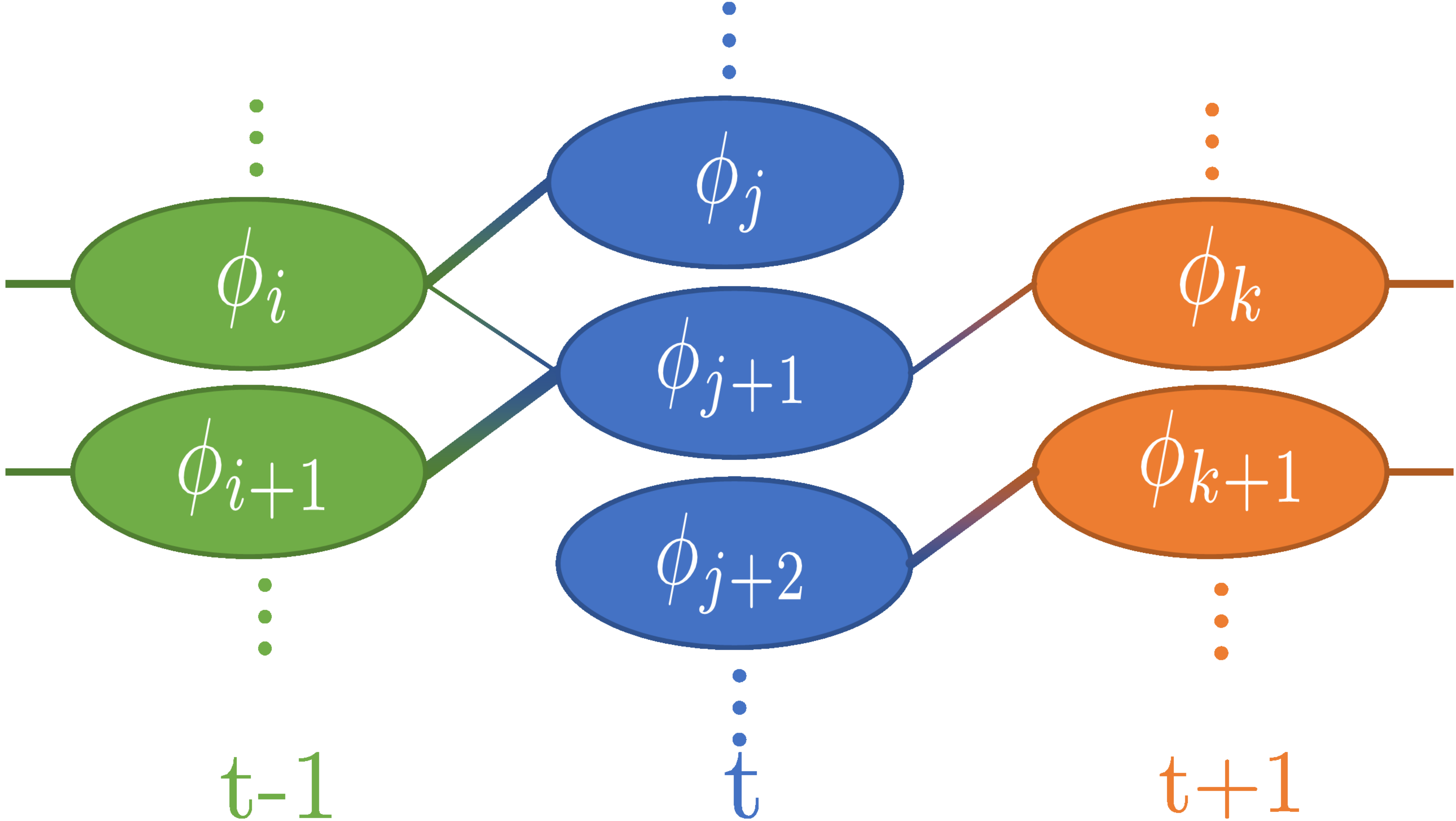}}
  \subfloat[Number of topics per epoch\label{fig:number-of-topics-epoch}]
     {\includegraphics[width=0.5\textwidth]{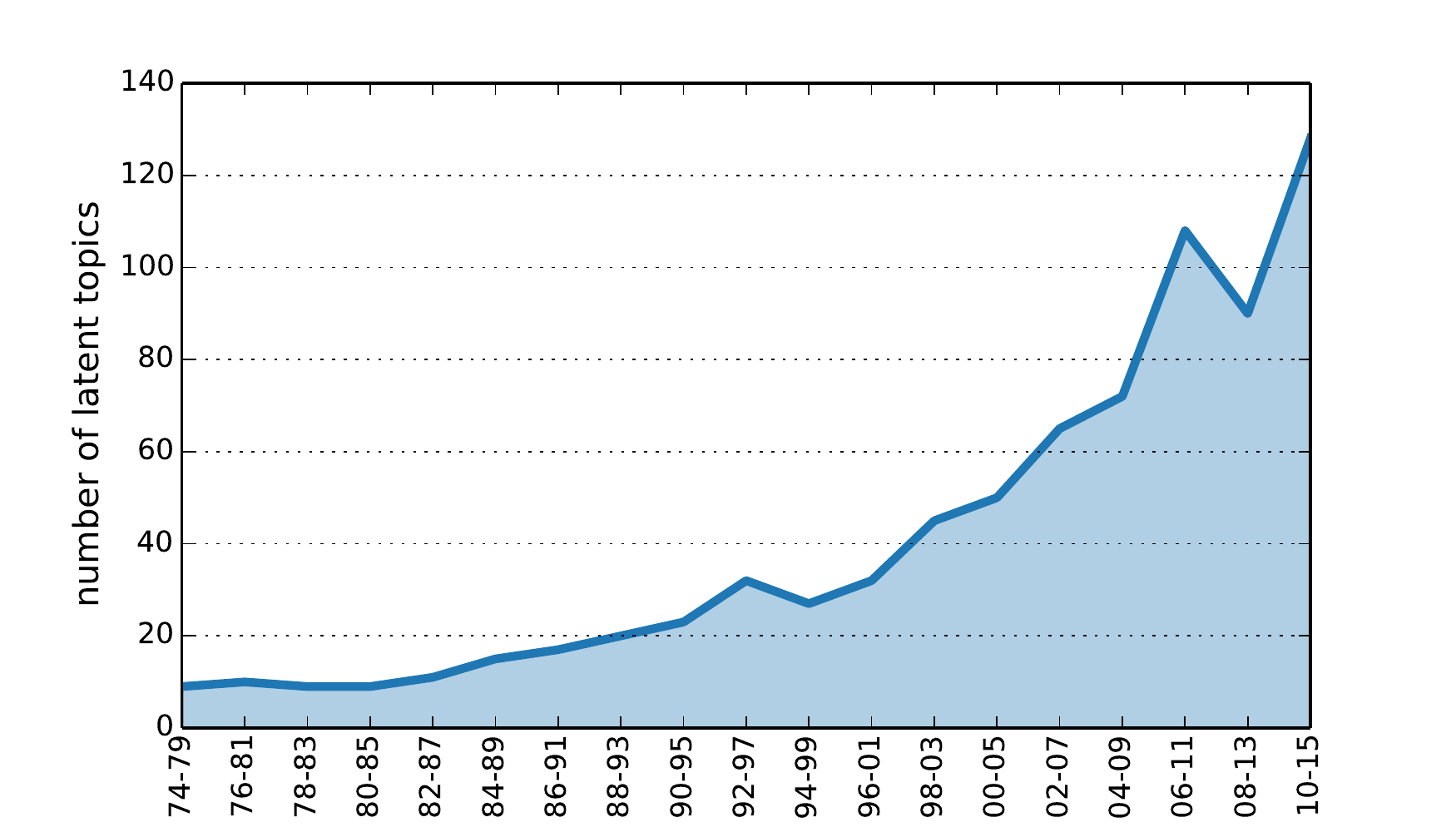}}
  \caption{(a) Conceptual illustration of the proposed similarity graph that models topic dynamics over time. A node corresponds to a topic in a specific epoch; edge weights are equal to the corresponding topic similarities. (b) As the document corpus grows so does the number of topics needed to model its latent structure. }
  \vspace{-10pt}
\end{figure}

\vspace{-4.5pt}
We divided the 42 year timespan of our data corpus into overlapping five year epochs, with a two year lag between consecutive epochs, resulting in 18 epochs in total. The topics of each epoch were then extracted as described in Section~\ref{sub:Modelling-Topics-Sequentially} and their dynamics inferred as per Section~\ref{sub:Modelling-Topics-Similarity}. The number of latent topics of different epoch is plotted in Figure~\ref{fig:number-of-topics-epoch}. Notice the exponential rise in the number of topics which mirrors the exponential increase in the number of publications over time in our dataset. This increasing interest in ASD can be illustrated by the observation that in 2013 there are five times as many publications as in 2000. For our inter-topic similarity described in Section~\ref{sub:Modelling-Topics-Similarity} we adopted the use of the well-known Jaccard similarity; this similarity measure was used to obtain all results reported in this section. Lastly, Gibbs sampling was used for HDP inference, implemented in Python~2.7, with hyperparameter resampling as described by Teh \textit{et al.}~\cite{teh_etal_2006_hierarchical}.

\begin{figure}
\vspace{-10pt}
  \centering
  \includegraphics[width=1\textwidth]{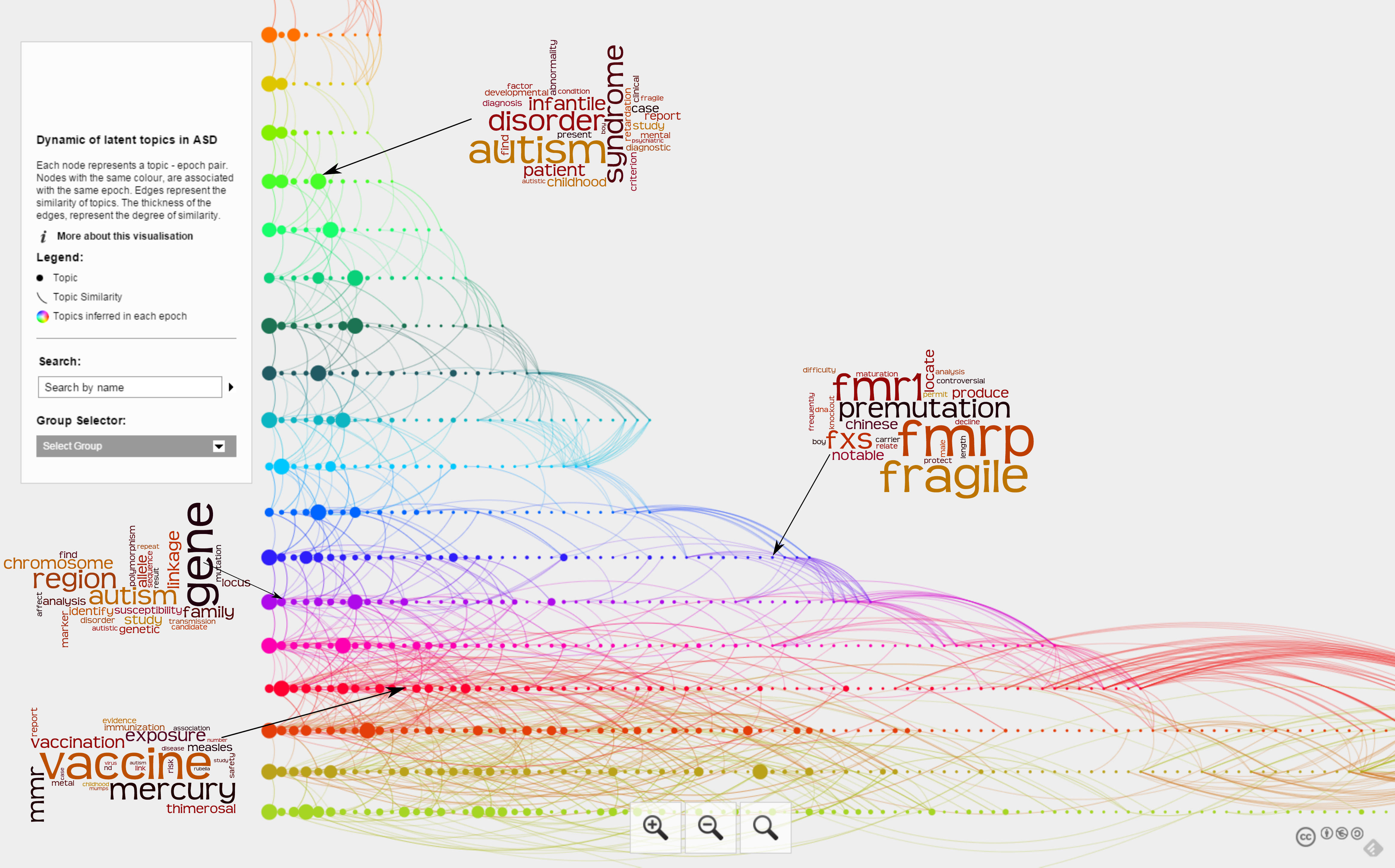}
  \caption{Interactive similarity graph analysis tool (see \url{www.goo.gl/Ws7V64}). Word clouds of a few topics are shown for illustration. Nodes and links between them represent respectively topics in particular epochs and their similarities.}
  \label{fig:topic_similarity_graph_result}
  \vspace{-10pt}
\end{figure}

\vspace{-4.5pt}\subsection{Case study 1: ASD and genetics\label{sub:Case-Study:genetics}}\vspace{-4.5pt}
While the exact aetiology of the ASD is still poorly understood, the existence of a significant genetic component is beyond doubt~\cite{Mile2011}. Work on understanding complex genetic factors affecting the development of autism, which possibly involve multiple genes which interact with each other and the environment, is a major theme of research and as such a good case study on which the usefulness of the proposed method can be illustrated.

\begin{figure}
\vspace{-10pt}
  \centering
  \includegraphics[width=0.95\textwidth]{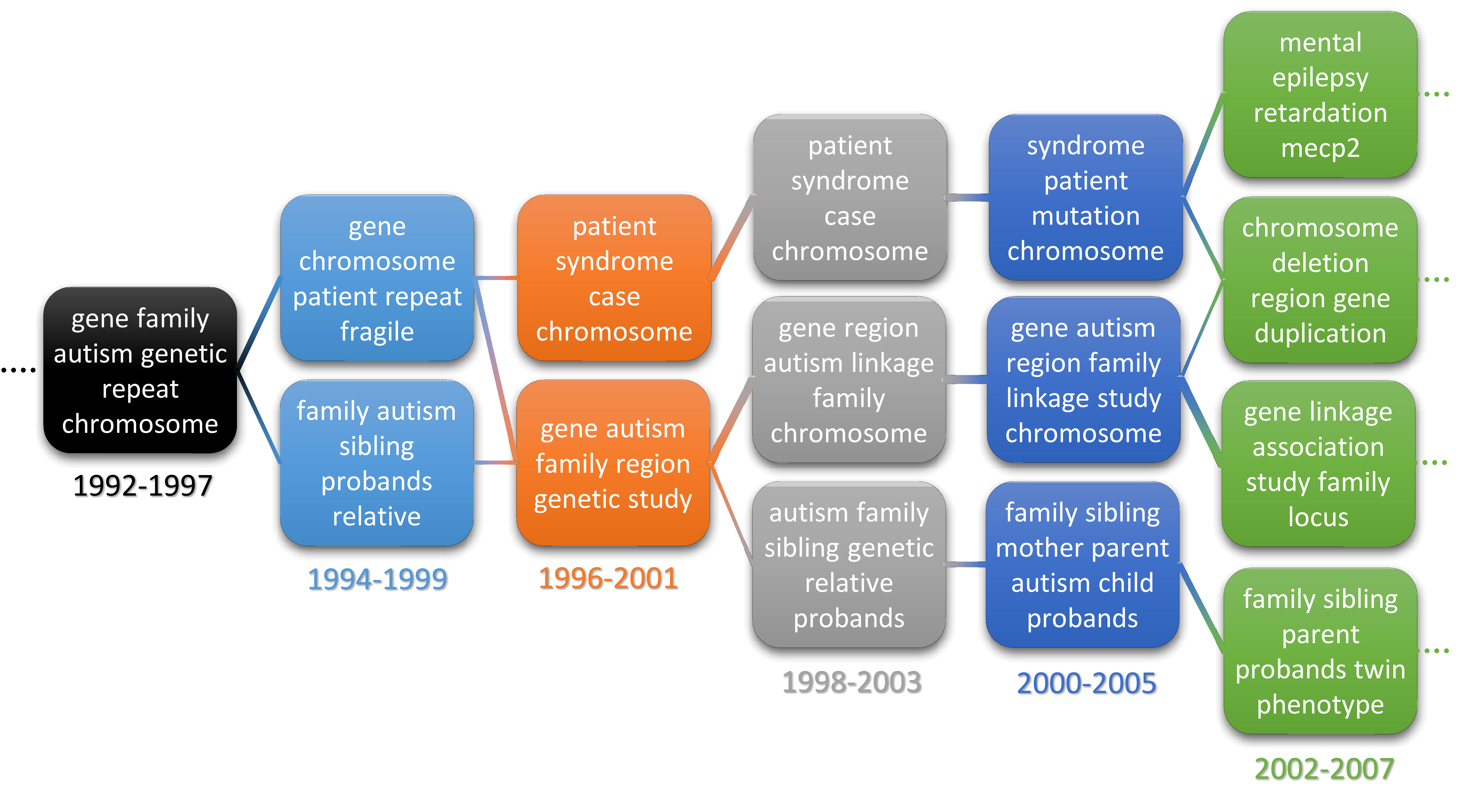}
  \caption{Dynamics of the topic most closely associated with the concept of ``genetics''. A few dominant words are shown for each topic (shaded boxes).}
  \label{fig:Dynamics-of-ASD}
  \vspace{-10pt}
\end{figure}

We started by identifying the topic of interest as that with the highest probability of the terms ``gene'' or ``genetic'' conditioned on the topic, and tracing it back in time to the epoch in which it originated. This led to the discovery of the relevant topic in the epoch spanning the period 1986--1991. Figure~\ref{fig:Dynamics-of-ASD} shows the evolution of this topic from 1992 revealed by our method (due to space constraints only the most significant parts of the similarity graph are shown; minor changes to the topic before 1992 are also omitted for clarity, as indicated by the dotted line in the figure). Each topic is labelled with its first few dominant terms. The following interpretation of our findings is readily apparent. Firstly, in the period 1992--1997, the topic is rather general in nature. Over time it evolves and splits into topics which concern more specific concepts (recall that such splitting of topics cannot be captured by any of the existing methods). For example by the epoch 2002--2007 the single original topic has evolved and split into four topics which concern:
\begin{itemize}
  \item the relationship between mutations in the gene \texttt{mecp2} (essential for normal functioning of neurone), and mental disorders and epilepsy  (it is estimated that one third of ASD individuals also have epilepsy),\\[-5pt]
  \item gene alternations, for example the duplication of \texttt{15q11--13} and deletion
of \texttt{16p11.2} both of which are associated with ASD,\\[-5pt]
  \item genetic linkage association analysis and heritability of autism, and\\[-5pt]
  \item observational work on autistic twins and probands with siblings on the spectrum.
\end{itemize}

Our framework also allows us to look `back' in time. For example, by examining the topics that the 1992 genetics topic originate from we discovered that the topic evolved from the early concept of ``infantile ASD''~\cite{kanner1946irrelevant}.

\vspace{-4.5pt}\subsection{Case study 2: ASD and vaccination\label{sub:Case-Study:vaccine}}\vspace{-4.5pt}
For our second case study we chose to examine research on the relationship between ASD development and vaccination. This subject has attracted much attention both in the research community, as well as in the media and the general public. The controversy was created with the publication of the work by Wakefield~\cite{WakeMurcAnth1998} which reported epidemiological findings linking MMR vaccination and the development of autism and colitis. Despite the full retraction of the article following the discovery that it was fraudulent, and numerous subsequent studies who failed to show the claimed link, a significant portion of the general public remains concerned with the issue.

\begin{figure}
  \vspace{-10pt}
  \centering
  \includegraphics[width=1\textwidth]{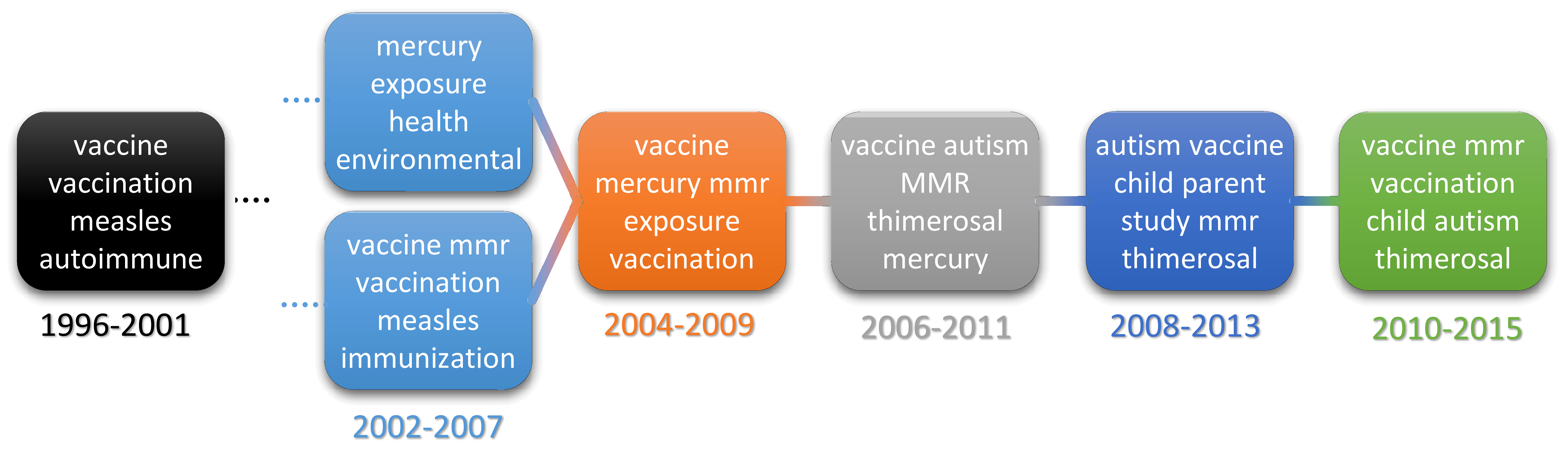}
  \caption{Dynamics of the topic most closely associated with the concept of ``vaccination''. Notwithstanding the rejection of any link between vaccination and autism, this topic remains active albeit in a form which evolved over time.}
  \label{fig:Dynamics-of-vaccination}
  \vspace{-10pt}
\end{figure}

As in the previous example, we begun by identifying the topic with the highest probability of the terms ``vaccine'' and ``vaccination'' conditioned on the topic, and tracing it back to the epoch in which it first emerged. Again, a single topic was readily identified, in the epoch spanning the period 1996--2001. Notice that this is consistent with the publication date of the first relevant publication by Wakefield~\cite{WakeMurcAnth1998}. The evolution of the topic is illustrated in Figure~\ref{fig:Dynamics-of-vaccination} in the same way as in the previous section. It can be seen that the original topic concerned the subjects initially brought to attention such as ``measles'', ``vaccine'', and ``autism''. In the subsequent epoch, when the original claim was still thought to have credibility, the topic evolves and splits into numerous others mirroring research directions taken by various researchers. Following this period and the revelations of its fraudulence, the topic assumes mainly single-threaded evolution, at times incorporating various originally separate ideas. For example observe the independent emergence of the term ``mercury''. Though initially unrelated to it this topic merges with the topic that concerns vaccination which can be explained by the widely publicized thiomersal (vaccine preservative) controversy (again note that such merging of topics cannot be captured by the existing methods). Although rejected by the medical community due to a lack of evidence, this topic can be seen as persisting to date.

\vspace{-4.5pt}\subsection{Topic browser}\vspace{-4.5pt}
A topic model can be seen as a dimensionality reduction framework that reduces documents into a topic space. This transformation of data can provide powerful insight and allow for the browsing of documents in a more subject-specific, semantic manner. For example by describing documents in the topic space, documents most related to a particular topic of interest can be readily identified and retrieved. To provide this functionality to the research community interested in ASD we used the framework described in this paper to model the entire literature corpus we collected, and built a website to facilitate free and ready use of our model and data. Researchers can use our online tool to browse topics, annotate them, and navigate through publications by topic. The website is available at \url{http://www.understandingautism.tk}.

\vspace{-4.5pt}\section{Conclusions\label{sec:Conclusions}}\vspace{-4.5pt}
We described a novel framework for temporal modelling of the topical structure of a longitudinal document corpus. Our approach consists of discretizing time into overlapping epochs, modelling the static topic structure within each epoch using an HDP, and tracking the evolution of topics over time using an inter-topic similarity measure. The resultant similarity graph captures relationships between topics in different epochs and allows for the automatic inference of the time of emergence and disappearance of topics, their evolution over time, merging and splitting. The power of the proposed general framework was demonstrated on the example of ASD-related medical literature. On two case studies which concern two important research issues in ASD literature we demonstrated that our method extracts meaningful topics and their temporal changes. A novel data corpus and free online tools are made freely available to researchers.

\bibliographystyle{splncs}
\bibliography{2015_PAKDD_paper1}

\end{document}